\documentclass[10pt,aps,prd,superscriptaddress,twocolumn]{revtex4-1}

\usepackage{hyperref}
\usepackage{amsmath}
\usepackage{amssymb}
\usepackage{graphics}
\usepackage{graphicx}
\usepackage{dcolumn} 

\newcommand{\pt}{\ensuremath{\mathrm{p}_{T}}}
\newcommand{\fb}{\ensuremath{\mathrm{fb}^{-1}}}
\newcommand{\ab}{\ensuremath{\mathrm{ab}^{-1}}}
\newcommand{\Eslash}{\mbox{$\rm E \kern-0.6em\slash$}}
\newcommand{\etmiss}{\mbox{$\rm \Eslash_{T}\!$}}

\newcommand{\GeV}{\text{GeV}}
\newcommand{\invfb}{\text{fb}^{-1}}
\newcommand{\ttbarXX}{\mbox{$t\bar{t}+\chi\bar\chi$}}

\begin{document}

\setlength{\pdfpageheight}{\paperheight}
\setlength{\pdfpagewidth}{\paperwidth}

\title{Prospects for collider searches for dark matter with heavy quarks}

\author{Giacomo Artoni}
\email[Email: ]{giacomo.artoni@cern.ch}
\affiliation{Martin A. Fisher School of Physics,
 Brandeis University, Waltham, MA 02453 }

\author{Tongyan Lin}
\email[Email: ]{tongyan@kicp.uchicago.edu}
\affiliation{Kavli Institute for Cosmological Physics,
                  University of Chicago, Chicago, IL 60637}

\author{Bj\"orn Penning}
\email[Email: ]{penning@cern.ch}
\affiliation{ \mbox{Fermilab, P.O.~Box 500, Batavia, IL 60510, USA} }
\affiliation{ Enrico Fermi Institute,
  University of Chicago, Chicago, IL 60637}

\author{Gabriella Sciolla}
\email[Email: ]{sciolla@cern.ch}

\author{Alessio Venturini}
\email[Email: ]{alessio.venturini@cern.ch}
\affiliation{Martin A. Fisher School of Physics,
 Brandeis University, Waltham, MA 02453 }
 
\begin{abstract}
  We present projections for future collider searches for dark matter
  produced in association with bottom or top quarks. Such production
  channels give rise to final states with missing transverse energy
  and one or more $b$-jets. Limits are given assuming an effective
  scalar operator coupling dark matter to quarks, where the dedicated
  analysis discussed here improves significantly over a generic
  monojet analysis. We give updated results for an anticipated
  high-luminosity LHC run at 14 TeV and for a 33 TeV hadron collider.
\end{abstract}

\date{\today}
\maketitle
 
\section{\label{sec:intro} Introduction}

The properties of dark matter (DM), and its connection to known particles
and physical laws, remain unknown. The WIMP hypothesis has guided the
quest to understand this fundamental problem; however, rapid
experimental progress continues to exclude historically popular models
of dark matter. In light of this uncertainty, more model-independent
searches for dark matter have gained significant traction. It is
important to be able to combine collider searches, direct
detection, and indirect detection in a complementary way, if we are to
determine that any signal seen in an experiment is indeed from
dark matter \cite{Bauer:2013ihz}.

In particular, the effective operator approach takes a simplifying
approach that allows one to relate the different experimental
signatures in a model-independent way. Here it is assumed that new
particles mediate interactions between dark matter and standard model
particles, but that these particles are too heavy to be produced in
experiments. Then the interaction can be described by a contact
operator \cite{Beltran:2010ww}, and it possible to classify the
operators in a systematic way. For each operator the relic abundance,
direct detection signal, and collider predictions depend on a single
parameter, $M_*$, and thus can be related simply.

The collider search for dark matter is a critical element of this
approach, and can provide the strongest constraints in cases where the
dark matter mass is below $\lesssim 10\ \GeV$ or when the operator has
suppressed direct detection signals.  The generic signature of dark
matter pair production in colliders is missing transverse energy from
the dark matter and energetic visible particles that are used to tag
the event.  For contact operators with couplings between dark matter
and quarks, the final state is a so-called ``monojet'' final state
consisting of one or two hard jets plus large missing transverse
energy from the dark matter. Experimental limits using monojet final
states have been published using 7 and 8 TeV LHC data
\cite{ATLAS:2012ky,ATLAS-CONF-2012-147,Chatrchyan:2012me,CMS-PAS-EXO-12-048}
for a variety of operators. In addition, similar final states with the
form $\chi \bar \chi + X$, where $\chi$ is the dark matter particle
and $X$ can be a photon, jet, or other particle, have been studied in
the context of effective operators
(e.g. \cite{Goodman:2010yf,Fox:2011pm,Petriello:2008pu,Fox:2011fx}).

In this note we focus on an effective scalar interaction between dark
matter and quarks described by the operator 
\begin{equation}
  {\cal O}= \sum_q \frac{m_q}{M_*^3}\bar{q}q\bar{\chi}\chi,
\end{equation}
where $M_*$ parameterizes the strength of the interactions.  The most
generic effective scalar interaction could include more complicated
couplings between different quark flavors, but this would lead to
flavor violating effects. An ansatz which automatically suppresses
these effects is minimal flavor violation (MFV)
\cite{D'Ambrosio:2002ex}, which then fixes the $m_q$ scaling of the
operator above. Because of the form of the interaction, couplings to
bottom and top quarks are significantly enhanced over those to light
quarks. This means that traditional monojet searches provide
relatively weak constraints on this scalar operator.

\begin{figure*}[tbh]
 \includegraphics[width=0.3\textwidth]{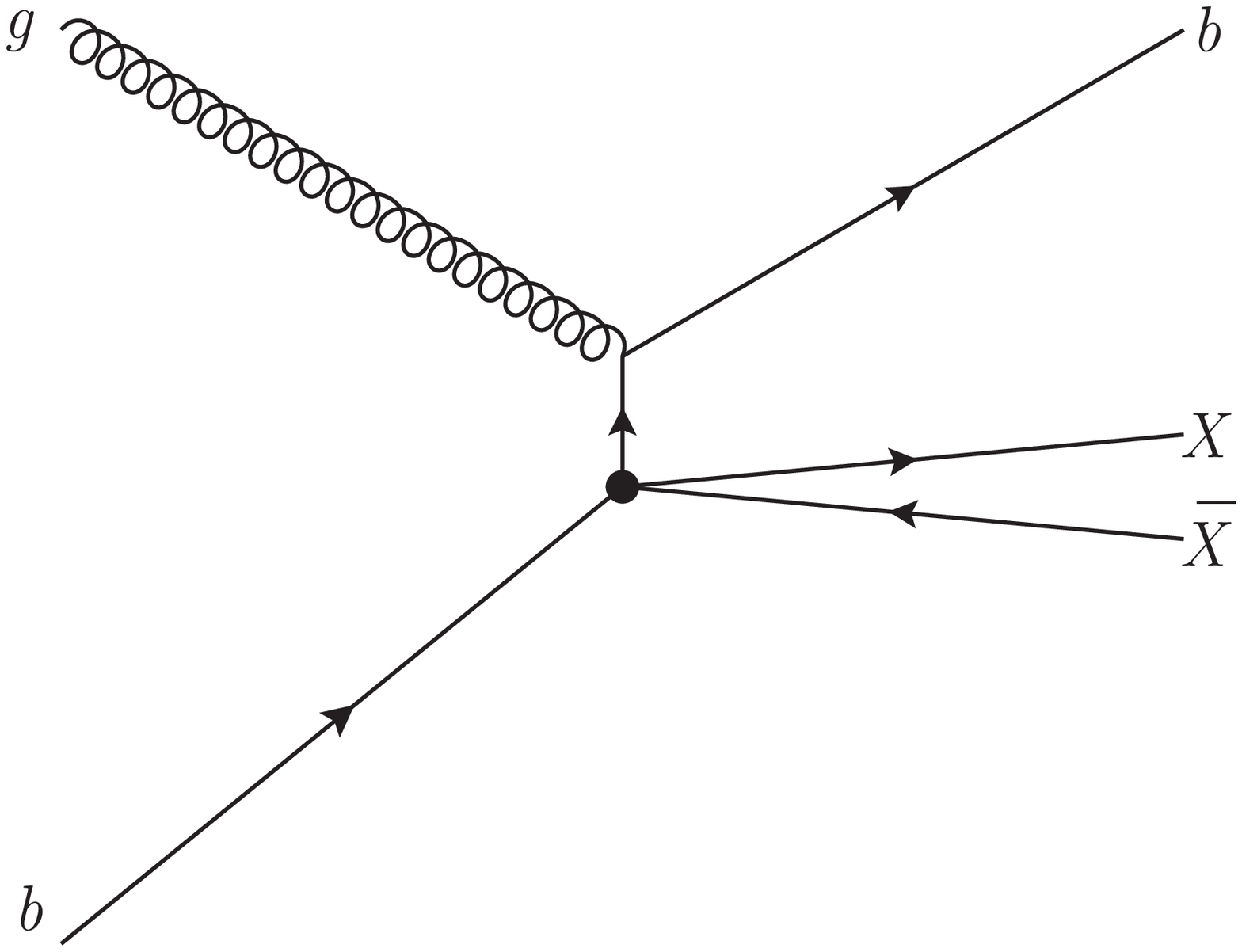}\hspace{.2cm}
 \includegraphics[width=0.3\textwidth]{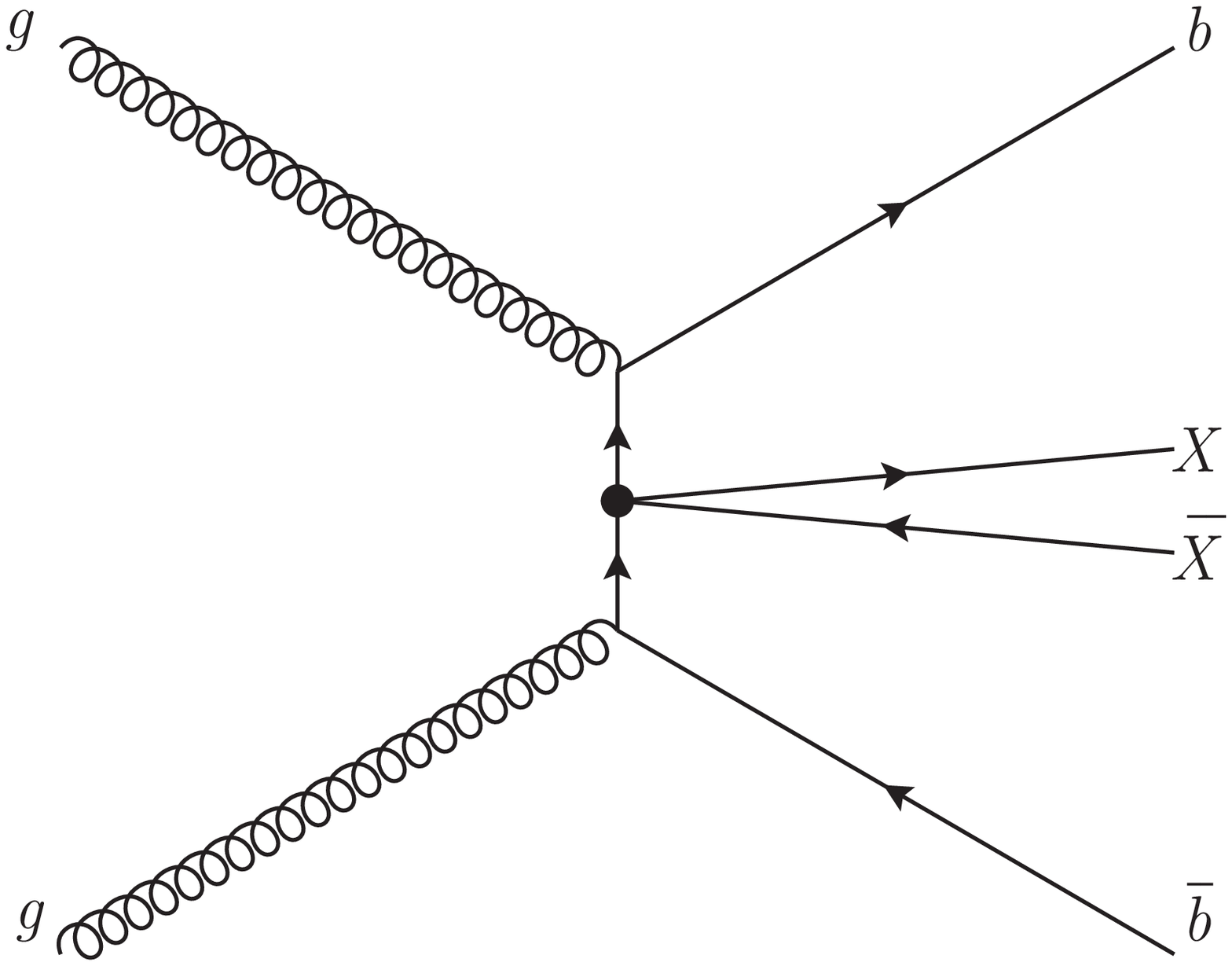}\hspace{.2cm}
 \includegraphics[width=0.3\textwidth]{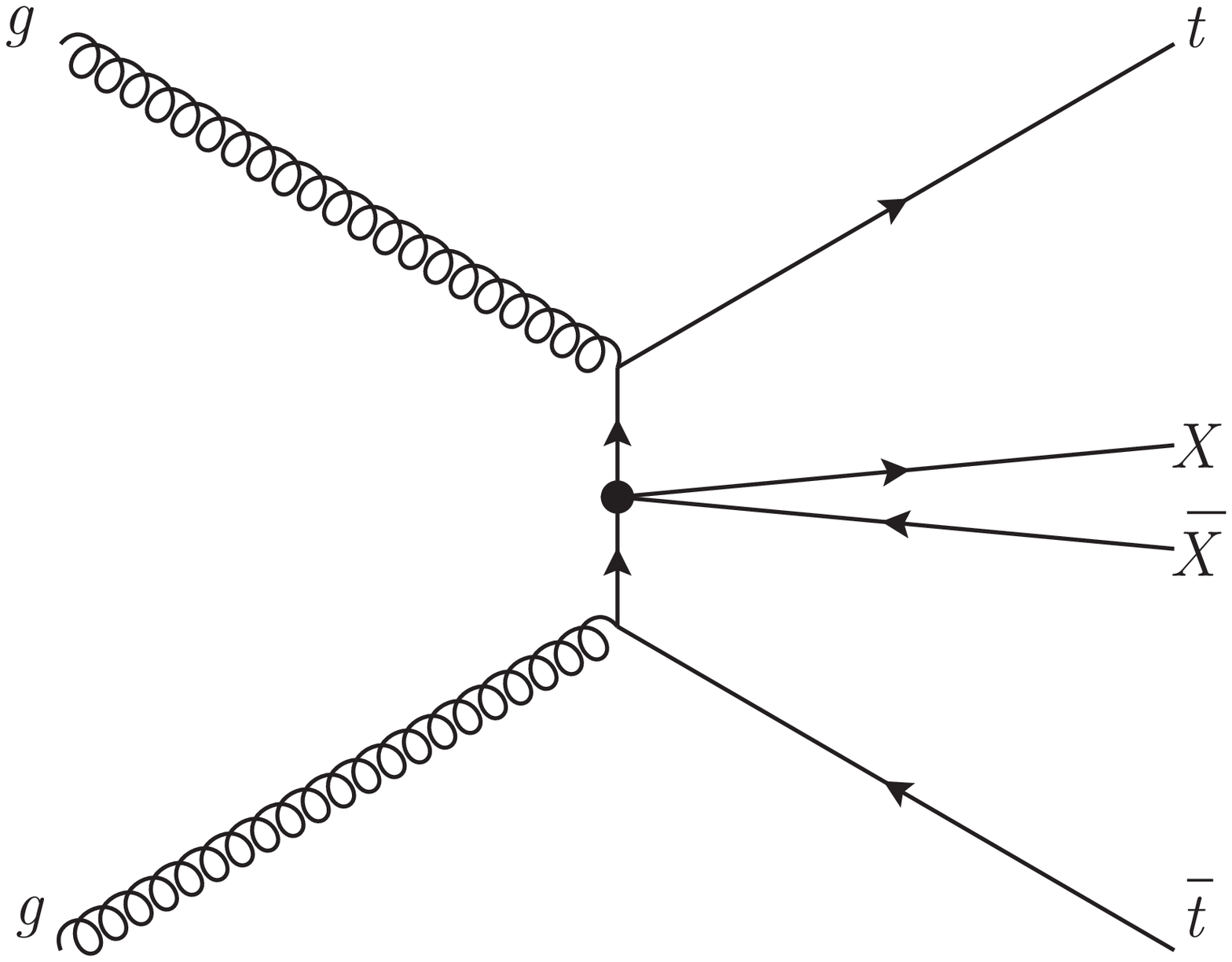}
\caption{\label{fig:diagrams} The dominant diagrams for dark matter
  plus heavy quark production.}
\end{figure*}

Despite the kinematic and PDF suppression for producing third
generation quarks, it was shown in \cite{Lin:2013sca} that for the
scalar operator, searches for $\chi \bar \chi + b$ and $\chi \bar \chi
+ t \bar t$ final states improve significantly over traditional
monojet searches. The production mechanism of dark matter plus heavy
quark modes at a hadron collider is shown in
Fig.~\ref{fig:diagrams}. In terms of limits on the direct detection
cross section, the improvement is up to three orders of
magnitude. Analyses of 8 TeV LHC data with these final states is
already underway. In this report we present expected limits for a 14
TeV LHC run with $300\ \invfb$ and $3\ \ab$ of data, and a 33 TeV $pp$
collider with $3\ \ab$ of data.

\section{\label{sec:simulations} Simulation}

We simulate signal events using MadGraph 5 \cite{Alwall:2011uj} and
PYTHIA \cite{Sjostrand:2006za} to model parton events and showering
with MLM matching. The signal (background) events are produced using a
MadGraph implementation provided from the work documented in
Ref.~\cite{Goodman:2010ku} (\cite{snowmassmadgraph}).  Cross sections
for dark matter in association with $b$-quarks are normalized using
MCFM Dark \cite{Fox:2012ru}, which calculates NLO corrections for dark
matter production processes.

Detector effects are simulated using the fast multipurpose detector
response simulation package DELPHES~3~\cite{deFavereau:2013fsa} in the
Snowmass Detector configuration~\cite{snowmassdetector}.  The
simulation includes a tracking system, embedded into a magnetic field,
calorimeters and a muon system with performances similar to that of
the Run~2 LHC detectors.  In the DELPHES simulation used, the
$b$-tagging efficiency is roughly $70\%$ with a mistag rate of $1\%$
for light quarks and $10\%$ for charm.

In this work, we have considered three  scenarios for future LHC datasets: 
\begin{enumerate} 
\item LHC Run 2 (R2-LHC): 300~\fb\            at 14~TeV with pileup of 50; 
\item High-luminosity LHC (HL-LHC): 3000~\fb\ at 14~TeV with pileup of 140; 
\item High-energy LHC (HE-LHC): 3000~\fb\     at 33~TeV with pileup of 140; 
\end{enumerate}

%
%

\section{ Analysis Strategy}

This analysis focuses on the so-called mono-$b$ signature, analogous to
the monojet signature but with the additional requirement of a
$b$-tag.  The search strategy is based on identifying the large
\etmiss{} signature from the DM pair recoiling against an energetic $b$
quark in the final state.  Because in the model the dark matter pair
production is mediated by heavy unobserved particles
we expect a significant \etmiss{} signature even for light dark matter
candidates.

For the mono-$b$ analysis, both DM production in association with
$b$-quarks and tops are important, although they differ qualitatively:
\begin{itemize}
  \item $bg \to \bar \chi \chi + b$, $gg \to \bar \chi \chi + b \bar
    b$: for these production modes, the jet multiplicity is low and
    most of the events have only one reconstructed $b$-jet; 
  \item $gg \to t\bar{t}+ \bar \chi \chi$: these final states have a
    higher probability of having two reconstructed $b$-jets and a
    harder $\etmiss$ spectrum. Although the jet multiplicity is high,
    the overall rate is large and thus this channel is important
    for a mono-$b$ analysis.
\end{itemize}

Events with at least one tagged b-jet of $\pt >$~50 GeV and
$\etmiss$>~100 GeV are considered in this study.  To suppress
backgrounds from $Z$+jets, $W$+jets and $t\overline{t}$ Standard Model
production, we veto events with leptons in the final state.

\begin{figure*}[tbh]
 \includegraphics[width=0.45\textwidth]{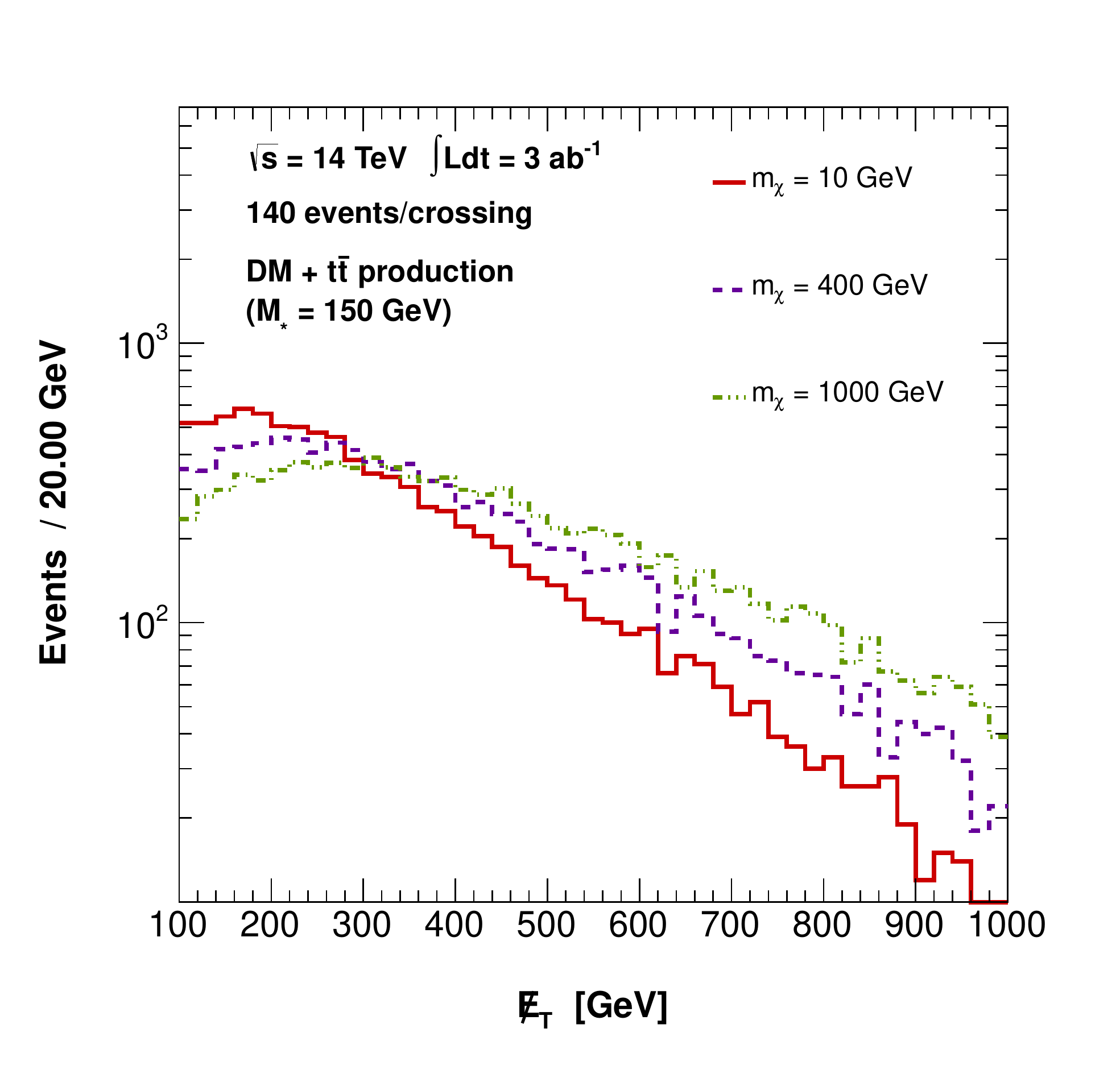}\hspace{.2cm}
 \includegraphics[width=0.45\textwidth]{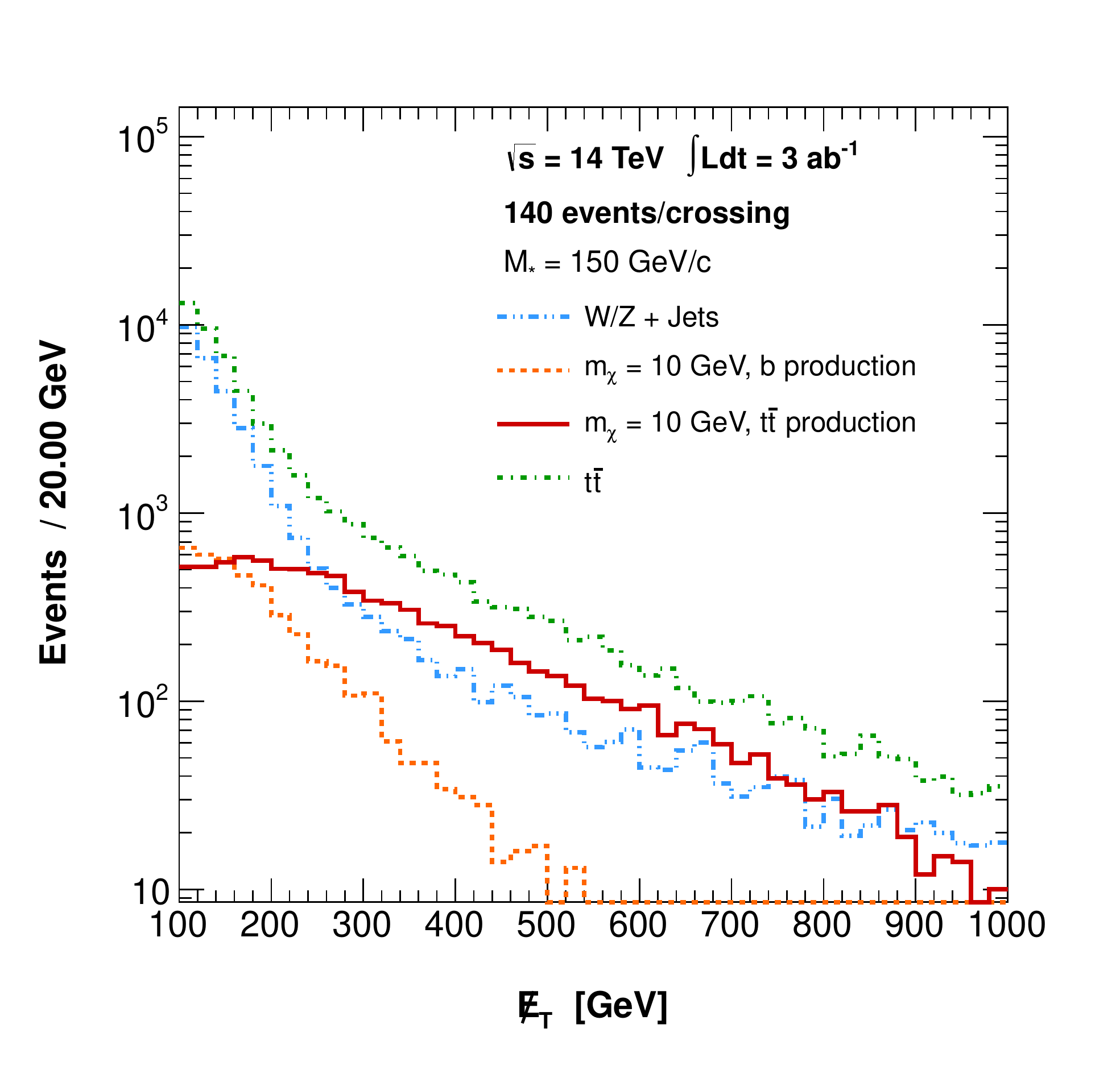}\hspace{.2cm}
 \caption{\label{fig:met14} 
Left: \etmiss~distribution of
  \ttbarXX{} production for different DM masses $m_\chi$ assuming 
  $\sqrt{s}=14~\textrm{TeV}$, pile up of 140 events per bunch crossing
  and an integrated luminosity of $3\ \ab$. 
Right: 
  \etmiss~distribution for direct $b$ and \ttbarXX{} production compared to
  backgrounds from $W/Z+\textrm{jets}$ and $t\bar{t}$ assuming 
  $\sqrt{s}=14~\textrm{TeV}$, pile up of 140 events per bunch crossing
  and an integrated luminosity of $3\ \ab$.  }
\end{figure*}

Figure~\ref{fig:met14} (left) shows the $\etmiss$ distribution obtained
in the case of top production for several different dark matter
masses and assuming $M_*$=150 GeV.  Figure~\ref{fig:met14} (right) shows
the same distribution for both direct $b$ production and top
production with a dark matter particle with a mass of 10 ~GeV.  The
same plot also shows the $\etmiss$ distribution for the dominant
background due to events with $W/Z + $jets in which a heavy flavor jet
is produced or a light quark jet is mistagged.  The distribution for
$t\bar t$ plus jets background is also shown.


For each dark matter mass assumption and luminosity scenario, 
the cut on $\etmiss$ cut is optimized by maximizing the sensitivity 
defined as  $S/\sqrt{S+B}$. 

\section{\label{sec:results} Results }

To obtain limit projections on $M_*$ and the DM-nucleon cross section
we produce signal samples for DM masses of $m_\chi$ = 1, 10, 50, 100,
200, 400, and 1000 GeV using both $b$ and $\bar t t$ production modes.
The analysis cuts were optimized for each DM mass value and for each
scenario discussed in section~\ref{sec:simulations}.
The projected sensitivity for DM plus heavy quark production was
calculated based on the total event rate.  The expected 90\% exclusion
limits on the dark matter$-$SM coupling, parameterized by the
suppression scale $M_*$, were computed for a given dark matter mass
$m_\chi$ by requiring $S/\sqrt{S+B} < 1.28$ for a one-sided Gaussian.

\begin{table}[t]
  \begin{tabular}{l|c|c|c}
    \hline 
    Process                &  Scenario 1      & Scenario 2   & Scenario 3 \\ 
    \hline
    \etmiss ~cut [GeV]      &   350     & 400  & 440 \\ 
    \hline
    \hline
    $Z/W +\textrm{jets}$  & 46141.2 &  301602.0  & 1778520.0    \\ \hline
    Top                   & 31759.8 &  203279.7  & 2193966.0       \\ \hline
    direct $b$              & 427.3   & 2825.943   & 89913.3 \\ \hline
    \ttbarXX              & 3927.7  & 29203.44    & 1699941 \\ \hline
  \end{tabular}
  \caption{Numbers of events 
    passing the selection requirements for the three considered
    scenarios. The \etmiss{} selections used are optimized for a DM
    particle of $m_\chi$ = 10 GeV.\label{tab:yields}}
\end{table}

\begin{figure*}[tb]
  \includegraphics[scale=0.4]{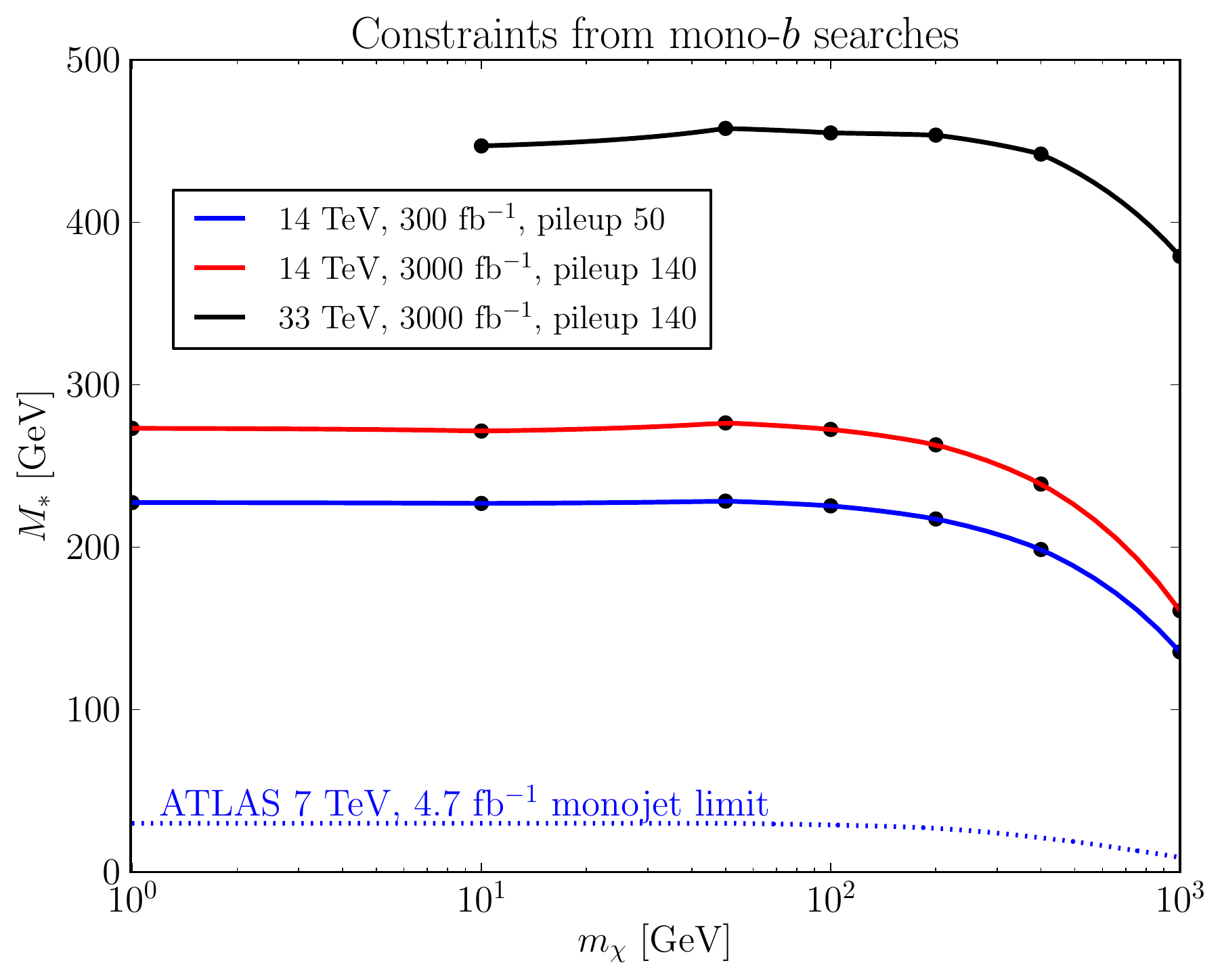}
  \includegraphics[scale=0.4]{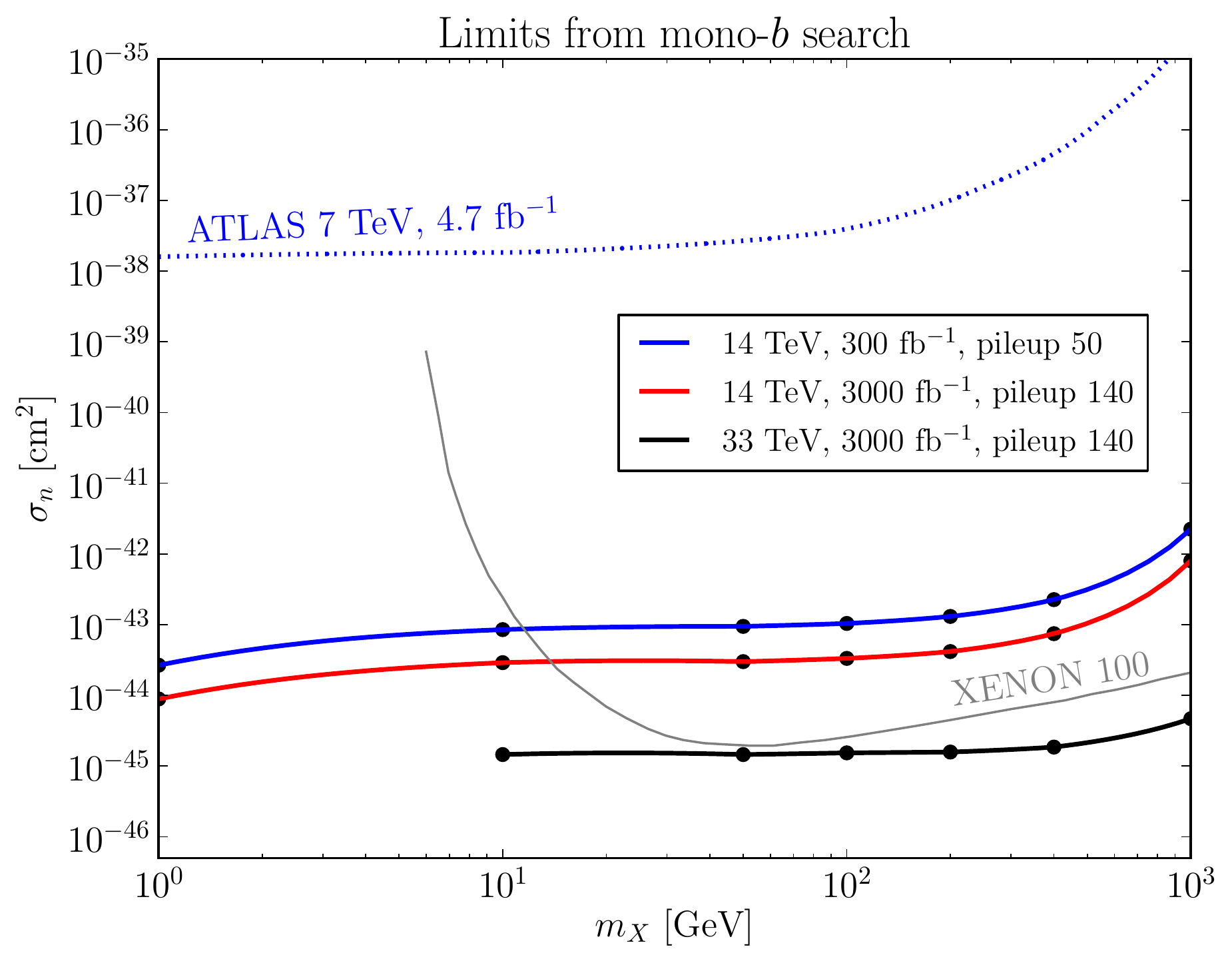}
\caption{\label{fig:combined_limits_mstar} Left: Expected 90\% CL
  limits on the scalar operator from a DM plus heavy quark search,
  including couplings to tops and bottoms. The limits for
  $\sqrt{s}=14~\textrm{and}~33~\textrm{TeV}$ with correspondingly
  adjusted \etmiss~ selections are shown. ATLAS 7 TeV limits come from
  \cite{ATLAS:2012ky}. Right: Corresponding constraints on the
  spin-independent nucleon scattering cross section, along with
  current XENON100 limits \cite{Aprile:2012nq}.}
\end{figure*}

Fig.~\ref{fig:combined_limits_mstar} shows the expected limits on the
$M_*$ for the three LHC scenarios mentioned above.  In mapping the
collider constraints to the direct detection plane, we relate $M_*$ to
the spin-independent nucleon scattering cross section for a scalar
operator:
\begin{align}
  \sigma_n  = \frac{(0.38 m_n)^2 \mu_{\chi n}^2}{\pi M_*^6}  \approx 2
  \times 10^{-38} \text{cm}^2 \left( \frac{30\ \GeV}{M_*} \right)^6.
\end{align}

%

\section{Discussion}

We have shown that limits on scalar interactions of dark matter with
quarks can be improved significantly by directly searching for final
states with $b$-quarks. The expected limits for three different
scenarios at the LHC were shown.



We show that limits on $M_*$ can improve by a factor of $\sim 7$ and
limits on $\sigma_n$ can improve by 5 orders of magnitude with 300
$\invfb$ of data from a 14 TeV run, compared to current monojet limits
using 7 TeV data from the LHC.
We also expect that limits can be further improved by taking advantage
of the shape differences between signal and background.

In addition to mono-b analysis discussed in this work, a dedicated
analysis of the $t\bar{t} + \bar \chi \chi$ final state will also
improve the limits. In \cite{Lin:2013sca} it was shown that $\sigma_n$
bounds were stronger by a factor of a few with respect to the mono-$b$
analysis. By probing both $b$ production and top production, we could
also gain insight into the flavor structure of dark matter couplings
with quarks.

Progress is underway both from the experimental side, where DM plus
heavy quark analyses are being applied to 8 TeV data, and from the
theoretical side, where work is being done to build UV complete models
and derive fully consistent constraints (for example
\cite{Dreiner:2013vla,Cotta:2013jna}). We anticipate this will lead to
some of the strongest complementary constraints on the direct
detection cross section, $\sigma_n$.

\bibliographystyle{unsrt}
\bibliography{monob} 

\end{document}